\documentclass[
  aps,
  prx,
  reprint,
  superscriptaddress,
  longbibliography
]{revtex4-2}

\usepackage[utf8]{inputenc}
\usepackage[T1]{fontenc}
\usepackage[english]{babel}

\usepackage{amsmath, amssymb, amsfonts, bm}
\usepackage{graphicx}
\usepackage{booktabs}
\usepackage{multirow}
\usepackage{array}
\usepackage{pifont}
\usepackage{enumitem}
\usepackage{url}
\usepackage{xcolor}
\usepackage{listings}
\usepackage[colorlinks=true, allcolors=blue]{hyperref}

\newcommand{\appsubhead}[1]{%
  \par\medskip
  \noindent\makebox[\linewidth][c]{\textit{#1}}%
  \par\smallskip
}

\begin{document}

\title{Latent Genetic Algorithm for Crystal Structure Prediction}

\author{Kaixin Zheng}
\affiliation{Key Laboratory of Computational Physical Sciences (Ministry of Education), Institute of Computational Physical Sciences, State Key Laboratory of Surface Physics, and Department of Physics, Fudan University, Shanghai 200433, China}

\author{Wanjian Yin}
\affiliation{College of Energy, Soochow Institute for Energy and Materials Innovations, Soochow University, Suzhou 215006, China}

\author{Hongyu Yu}
\email[Contact author: ]{hongyuyu20@fudan.edu.cn}
\affiliation{Key Laboratory of Computational Physical Sciences (Ministry of Education), Institute of Computational Physical Sciences, State Key Laboratory of Surface Physics, and Department of Physics, Fudan University, Shanghai 200433, China}

\author{Hongjun Xiang}
\email[Contact author: ]{hxiang@fudan.edu.cn}
\affiliation{Key Laboratory of Computational Physical Sciences (Ministry of Education), Institute of Computational Physical Sciences, State Key Laboratory of Surface Physics, and Department of Physics, Fudan University, Shanghai 200433, China}

\begin{abstract}

Predicting crystal structures requires navigating rugged energy landscapes in which favorable local motifs must be inherited across candidates with incompatible cells, densities, and symmetries. Conventional real-space crossover often destroys these motifs when parent structures are geometrically mismatched. Here we show that latent representations learned by pretrained universal interatomic potentials can serve as continuous evolutionary coordinates for crystal structure prediction. In the Latent Genetic Algorithm (LGA), offspring are generated by inverse optimization of atomic positions and lattice vectors to match a target latent representation, which is constructed via interpolation of the parent latent vectors. LGA suppresses high-energy and short-contact offspring, increases the HfO$_2$ ground-state recovery rate from 20--35\% to 60--95\%, and enables a unified variable-supercell search over 16 perovskites with a nearly tenfold reduction in search cost. Applied to \((\mathrm{PbTiO}_3)_n/(\mathrm{PbZrO}_3)_n\) superlattices, LGA reveals \(\sqrt{2}\times3\sqrt{2}\times1\) long-period ground-state structures characterized by a common in-plane finite-\(q\) modulation \(q_\parallel=(1/6,1/6)\) and layer-coupled sidebands. To our knowledge, this in-plane periodicity has not been reported in any related oxide perovskite superlattice studies. Altogether, LGA offers a powerful representation-guided paradigm for ground-state structure prediction and provides a practical, decoder-free route toward materials inverse design.

\end{abstract}

\maketitle

\section{Introduction}

Determining crystal structures is essential for understanding material properties and designing functional materials. Although experimental techniques, such as X-ray diffraction, remain central to structural characterization, they can be limited by synthesis conditions, sample quality, and the stochastic nature of crystal growth. With advances in computational methods and high-performance computing, crystal structure prediction (CSP)~\cite{CSP1,CSP2} has become an effective tool for materials discovery, enabling the identification of thermodynamically stable or metastable phases from first principles with minimal prior information. By exploring highly non-convex potential energy surfaces (PESs), many CSP algorithms~\cite{AGA,EVO,MAGUS1,PASP,Xtalopt,AIRSS,CALYPSO,Hopping,USPEX1,USPEX2,USPEX3} have achieved broad success across materials systems~\cite{Case-Caylpso,Case-AGA,Case-EVO,Case-MAGUS,Case-PASP1,Case-PASP2,Case-Xtal}.

Among these techniques, genetic algorithms (GAs)~\cite{GA} are widely used heuristic strategies. Inspired by biological evolution, GAs maintain a diverse population of candidate structures and iteratively evolve them toward low-energy minima on the PES. Despite their success, conventional GAs are often limited by real-space crossover operators. Widely used operations such as cut-and-splice directly manipulate fractional coordinates and lattice vectors. For structurally diverse parent populations, where lattice constants or atomic densities can differ substantially, splicing incompatible structures may break the local atomic environment, which is critical for structural stability. The resulting high-energy or invalid offspring increase the cost of subsequent DFT relaxations and can break the inheritance of favorable local structural motifs. This limitation reflects a mismatch between the quantities manipulated by conventional crossover operators and the physical information that should be inherited.

\begin{figure*}[!ht]
    \centering
    \includegraphics[width=\linewidth]{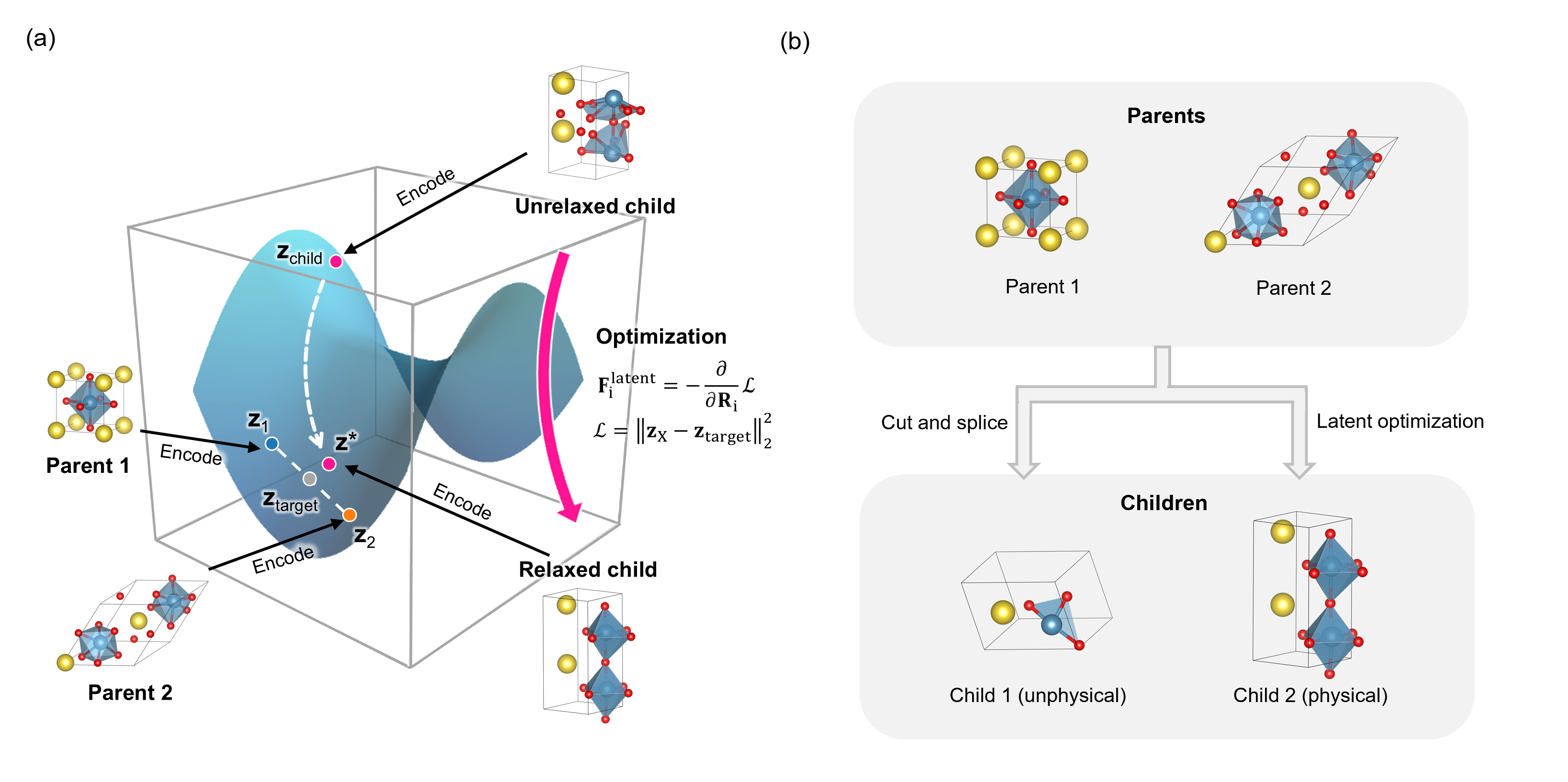}
    
    \caption{Latent-space crossover as representation-guided structural evolution.
(a) Schematic illustration of latent-space crossover. While conventional real-space crossover combines atomic coordinates and lattice fragments directly, LGA instead encodes each parent structure using the local-environment representation learned by a pretrained interatomic potential, interpolates the resulting global latent vectors to define \(z_{\mathrm{target}}\), and obtains an offspring by optimizing atomic positions and lattice vectors to match this target representation.
(b) Comparison between real-space cut-and-splice and latent-space optimization for the same parent pair. Cut-and-splice produces a geometrically incompatible child with disrupted coordination environments and short contacts, whereas latent optimization yields a physically plausible offspring that preserves local-environment information inherited from the parents.}
    \label{fig:schematic}
\end{figure*}

Pretrained universal interatomic potentials~\cite{Behler,CHGNET,M3GNET,MACE,Mattersim,Sevennet,UPET, UPET-MAD, Seven-omni} encode local atomic environments into high-dimensional latent vectors for subsequent energy and force prediction, offering an alternative representation of atomic structures. This suggests a different route for offspring generation, where crossover is performed within the latent space rather than real space. In this work, we introduce the Latent Genetic Algorithm (LGA), a representation-guided framework that uses latent atomic representations encoded by pretrained universal interatomic potentials as evolutionary coordinates. By leveraging the end-to-end differentiability of potentials, LGA performs smooth interpolation on the latent vectors of parent structures and directly optimizes for a new structure whose latent embedding matches the interpolated target. This decoder-free formulation converts crossover into a physically constrained inverse problem and avoids the structural failures associated with direct coordinate splicing. 

Our paper is organized as follows. Sec. \ref{mechanism} introduces the Latent Genetic Algorithm (LGA) mechanism, detailing how structural crossover is performed in latent space. Sec. \ref{validation} uses $\text{HfO}_2$ as a diagnostic system, demonstrating that latent-space crossover provides smoother structural interpolation and suppresses high-energy, short-contact offspring. Sec. \ref{benchmarks} evaluates the robustness and efficiency of LGA in full evolutionary searches, where it consistently improves ground-state recovery for $\text{HfO}_2$ across multiple encoders and reduces the search cost for perovskite systems by nearly tenfold. Sec. \ref{superlattice} focuses on physical discovery by extending LGA to $(\mathrm{PbTiO}_3)_n/(\mathrm{PbZrO}_3)_n$ superlattices, revealing unusual long-period structures with a thickness-dependent finite-$q$ distortion channel. Finally, we summarize our findings and discuss future applications in Sec. \ref{discussion}.

%\section{Results}

\section{Latent-space crossover mechanism} \label{mechanism}

As depicted in Fig.~\ref{fig:schematic}(a), the Latent Genetic Algorithm (LGA) recasts structural crossover as interpolation in a learned representation of atomic environments. Instead of splicing fractional coordinates and lattice fragments directly, LGA encodes parent structures with a pretrained universal interatomic potential, interpolates their latent vectors, and obtains an offspring by inverse optimization of atomic positions and lattice vectors. This procedure transfers structural information through a chemically informed continuous space, reducing the risk of breaking local coordination motifs or collective periodic distortions when parent cells are geometrically incompatible.

Our construction allows structures with different atom numbers and lattice geometries to be compared in a common representation space. Formally, let $\mathcal{S}$ denote the space of crystal structures. We define a composite mapping function $\Psi: \mathcal{S} \rightarrow \mathbb{R}^d$ that projects any structure $X \in \mathcal{S}$ containing $N$ atoms into a continuous $d$-dimensional latent representation space. First, the pretrained atomic encoder, denoted as $\phi(\cdot)$, maps the local environment of each atom $i$ to an embedding vector $\mathbf{h}_i \in \mathbb{R}^d$. Subsequently, to construct a size-invariant global representation $\mathbf{z}_X$ compatible with a variable number of atoms, we apply a global mean pooling operation:
\begin{equation}\label{Eq1}
    \mathbf{z}_X = \Psi(X) = \frac{1}{N} \sum_{i=1}^{N} \phi(\text{env}_i)
\end{equation}
The resulting vector $\mathbf{z}_X$ resides in a continuous Euclidean space $\mathbb{R}^d$. This vector-space representation enables linear operations for structural evolution.

During the crossover event between two parent structures $X_1$ and $X_2$, a target latent vector $\mathbf{z}_{\text{target}}$ is generated via weighted interpolation:
\begin{equation}
    \mathbf{z}_{\text{target}} = \omega_{1}\mathbf{z}_{1} + \omega_{2}\mathbf{z}_{2}
\end{equation}
where the weights $\omega_{1}$ and $\omega_{2}$ determine the hereditary contribution of each parent (specific strategies for determining these weights are detailed in Appendix~\ref{app:latent}). To isolate the effect of the latent-space crossover operator, all benchmarks in this study use the uniform weighting strategy, \(\omega_1=\omega_2=0.5\).

Finally, offspring generation is formulated as an inverse optimization task that seeks a structure whose model-encoded latent representation matches the interpolated target vector. We initialize a candidate structure with a random space group and optimize its atomic coordinates $\mathbf{R}$ and lattice parameters $\mathbf{H}$ to minimize its distance to the target vector in the latent space. The optimal offspring $X^*$ is obtained by solving:
\begin{equation}
    X^* = \operatorname*{arg\,min}_{X(\mathbf{R}, \mathbf{H}) \in \mathcal{S}} \mathcal{L}(\mathbf{R}, \mathbf{H}) \quad \text{where} \quad \mathcal{L} = \| \Psi(X) - \mathbf{z}_{\text{target}} \|_2^2
\end{equation}

The inverse optimization uses the differentiability of the pretrained potential with respect to atomic coordinates and lattice degrees of freedom. The negative gradients of the latent loss define force- and stress-like descent directions:

\begin{align}
    \mathbf{F}_i^{\text{latent}} &= -\nabla_{\mathbf{R}_i} \mathcal{L} \quad (\text{Virtual Force}) \\
    \mathbf{\sigma}^{\text{latent}} &= -\frac{\partial \mathcal{L}}{\partial \mathbf{H}} \quad (\text{Virtual Stress})
\end{align}

During optimization, \(F^{\mathrm{latent}}_i\) acts on the atomic coordinates, while the virtual stress \(\sigma^{\mathrm{latent}}\) acts on the lattice degrees of freedom. Together, they guide a randomly initialized candidate toward a structure whose latent-space representation matches the interpolated target. 

\begin{figure}[t!]
    \centering
    \includegraphics[width=\linewidth]{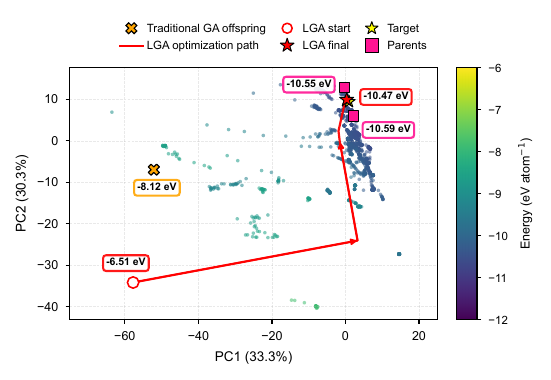}
    \caption{Representative visualization of MACE latent representations and latent optimization.
    The background scatter plot shows the PCA projection of global MACE latent representations for 100 HfO$_2$ relaxation trajectories, colored by their VASP energies. Symbols denote the parent structures, initial LGA offspring, optimization trajectory, interpolated target vector, final LGA offspring, and conventional GA offspring. For brevity, all energy labels are given per atom. Visualizations of the key structures are provided in the Supplemental Material.}
    \label{fig:pca}
\end{figure}

\begin{figure*}[htbp]
    \centering
    \includegraphics[width=0.9\linewidth]{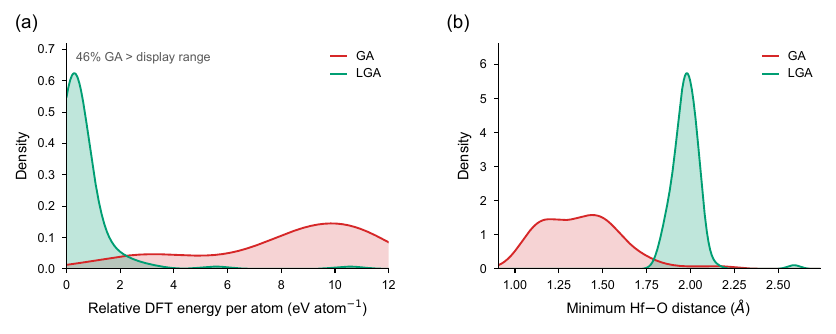}
    \caption{LGA suppresses offspring with high energies or short contacts.
    (a) Kernel density estimation (KDE) of the relative static DFT energies per atom for 100 offspring generated by conventional GA (red) and LGA (green). Relative energies are reported with respect to the lowest-energy structure in the comparison set. The annotation denotes the fraction of GA offspring outside the displayed energy range. (b) KDE of the minimum Hf--O distances in the same offspring structures. Conventional GA generates a broad distribution with many short Hf–O contacts, whereas LGA produces a narrower distribution centered near typical Hf–O coordination distances.}
    \label{fig:energy_dist}
\end{figure*}

\section{Validation of the latent-space crossover operator} \label{validation}

We next tested, at the level of individual crossover events, whether a pretrained representation reduces the structural failures produced by real-space crossover. Because MACE is used as the default latent encoder in the following benchmark analyses, we first use its representation as a representative diagnostic case. We encoded the relaxation trajectories of 100 HfO$_2$ configurations using MACE and projected the resulting global representations with principal component analysis (PCA).

As shown in Fig.~\ref{fig:pca}, the MACE projection shows a partial organization by energy: lower-energy configurations are enriched in a relatively compact region, whereas high-energy configurations are more broadly distributed. Starting from the same pair of parent structures ($-10.55$ eV atom$^{-1}$ and $-10.59$ eV atom$^{-1}$), while the conventional GA offspring  ($-8.12$ eV atom$^{-1}$) lies away from this low-energy-enriched region, latent-space crossover moves a randomly initialized candidate ($-6.51$ eV atom$^{-1}$) toward the interpolated parent target and yields a substantially lower-energy offspring ($-10.47$ eV atom$^{-1}$) after optimization (see Supplemental Material for visualizations of the key structures). This representative trajectory suggests that the MACE latent representation provides a smoother coordinate system for inheriting structural information than direct real-space crossover.

We then quantified whether this representation-guided crossover improves offspring quality. For each method, 100 offspring were generated from the same HfO$_2$ parent pool. Their static DFT energies were reported relative to the lowest-energy structure in the comparison set, and the minimum Hf--O distance was extracted as a local geometric indicator of short-contact artifacts. As shown in Fig.~\ref{fig:energy_dist}(a), conventional GA generates a broad high-energy distribution, with 46\% of GA offspring lying outside the displayed energy range. By contrast, LGA offspring are concentrated in a narrow low-energy regime. The minimum-distance distributions in Fig.~\ref{fig:energy_dist}(b) reveal the same trend, showing that conventional GA produces many offspring with unusually short Hf--O contacts, while LGA yields a much narrower distribution centered near typical coordination distances. At the operator level, these results indicate that latent-space crossover suppresses both high-energy candidates and geometrically unfavorable local environments, thereby generating offspring that are more suitable for subsequent evolutionary search.

\section{Algorithmic benchmarks} \label{benchmarks}

\subsection{Random-Based LGA search in HfO$_2$}

The Random-Based LGA benchmark in HfO$_2$ tests whether the mechanistic advantage observed for individual offspring persists in an unconstrained global evolutionary search. We evaluated LGA using four pretrained universal interatomic potentials as latent encoders: MACE~\cite{MACE}, MatterSim~\cite{Mattersim}, SevenNet~\cite{Seven-omni}, and UPET~\cite{UPET-MAD}. We employed the Random-Based LGA framework—detailed in Appendix~\ref{app:workflows} and implemented within the \textsc{PASP}~\cite{PASP} package—which initializes populations with symmetry-constrained random configurations. The search spans structures containing 3--24 atoms and requires the algorithm to recover the known 12-atom monoclinic ground-state primitive cell.

To establish a widely used full-search baseline, we compared the performance of LGA against the \textsc{USPEX}~\cite{USPEX1,USPEX2,USPEX3} code. In this HfO$_2$ benchmark, the pretrained potentials function exclusively as latent encoders during crossover, while all subsequent geometric relaxations, fitness rankings, and success assessments are performed uniformly via VASP~\cite{VASP1,VASP2,VASP3,VASP4}. For a fair comparison, both USPEX and Random-Based LGA were tested at three crossover probabilities, \(P_c \in \{0.25,0.50,0.75\}\). All searches shared a population size of 20 and evolved for 10 generations under matched VASP relaxation protocols (detailed in Appendix~\ref{app:computational}). As shown in Table~\ref{tab:success}, LGA consistently yields higher recovery rates than the conventional GA baseline across all tested pretrained potentials and $P_c$ values. These results connect the operator-level behavior in Fig.~\ref{fig:energy_dist} to a full evolutionary search comparison and show that the benefit of representation-guided crossover is not restricted to a single pretrained model architecture. 

\begin{table}[htbp]
    \centering
    \caption{LGA improves the success rate for HfO$_2$ structure search.
    Success rates are reported for recovering the 12-atom monoclinic ground-state primitive cell from a search space containing 3--24 atoms. Conventional real-space GA searches were performed using USPEX. Random-Based LGA searches used MACE, MatterSim, SevenNet, or UPET as the latent encoder. Values in parentheses indicate the number of successful runs out of 20 independent runs.}
    \label{tab:success}
    \begin{ruledtabular}
\begin{tabular}{lccc}
Method & \multicolumn{3}{c}{Success rate} \\
 & $P_c = 0.25$ & $P_c = 0.50$ & $P_c = 0.75$ \\
\colrule
GA (USPEX)      & 35\% (7/20)   & 20\% (4/20)   & 30\% (6/20) \\
LGA (MACE)      & 90\% (18/20)  & 85\% (17/20)  & 95\% (19/20) \\
LGA (MatterSim) & 80\% (16/20)  & 65\% (13/20)  & 70\% (14/20) \\
LGA (SevenNet)  & 80\% (16/20)  & 90\% (18/20)  & 95\% (19/20) \\
LGA (UPET)      & 60\% (12/20)  & 85\% (17/20)  & 80\% (16/20) \\
\end{tabular}
\end{ruledtabular}
\end{table}

\begin{figure*}[t]
    \centering
    \includegraphics[width=0.9\textwidth]{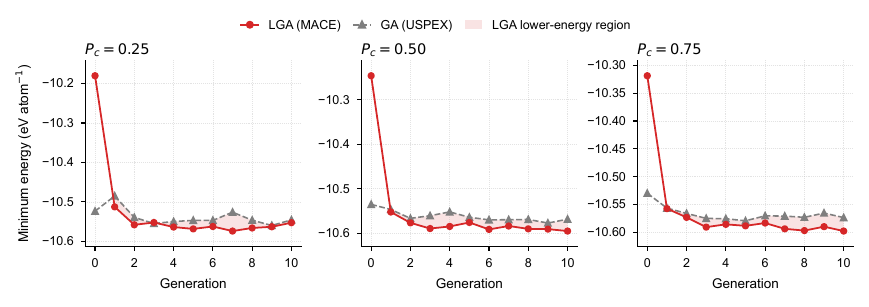}
    \caption{\label{fig:min_e}LGA consistently identifies lower-energy heredity offspring during evolution.
The minimum potential energy per atom is tracked across generations for the traditional GA (USPEX) and LGA (MACE) at different crossover probabilities \(P_c\). All trajectories are averaged over 20 independent runs and include only structures generated by heredity operators. The lower LGA (MACE) curves indicate that latent-space crossover improves the energetic quality of offspring throughout the evolutionary process.}
\end{figure*}

Beyond the final success rates, we also analyzed the evolutionary dynamics of the populations. As detailed in Fig.~\ref{fig:min_e}, we tracked the generation-wise evolution of minimum atomic energies, averaged over 20 independent runs for both GA (USPEX) and LGA (MACE). To strictly benchmark the crossover efficiency, this analysis focuses exclusively on offspring generated via heredity operators. The LGA consistently identifies deeper energy minima from generation 1 onward, across all crossover probabilities. This observation aligns with the superior success rates shown in Table~\ref{tab:success}, indicating that LGA is more effective at escaping local minima and refining candidates toward the true ground state.

\subsection{Perturbation-Based LGA search in perovskites} 

The Perturbation-Based LGA benchmark tests a different search setting, where candidate structures are generated as distortions of a high-symmetry aristotype while the relevant supercell periodicity is not known in advance. This setting is physically natural for perovskites, whose low-energy phases often emerge from symmetry-lowering distortions of the cubic reference structure. The conventional perturbation-based GA implemented in \textsc{PASP}, however, is usually confined to fixed supercells. When the correct periodicity is unknown, one must repeat the search over many candidate supercells, and direct real-space crossover between inequivalent supercells does not naturally preserve the perturbative character of the distortions. To address this limitation, we developed the Perturbation-Based LGA (described in Appendix~\ref{app:workflows}). By combining Hermite Normal Forms (HNFs) for discrete supercell topology with latent-space optimization for internal distortions, this formulation treats periodicity and internal structure as coupled evolutionary degrees of freedom in a single search.

We benchmarked this approach across 16 distinct perovskite systems, comprising 12 $A^{2+}B^{4+}\mathrm{O}_3$ compounds ($A=\mathrm{Ca}$, $\mathrm{Pb}$, $\mathrm{Ba}$, $\mathrm{Sr}$;\ $B=\mathrm{Ti}$, $\mathrm{Zr}$, $\mathrm{Hf}$) and 4 $A^{+}B^{5+}\mathrm{O}_3$ compounds ($A = \mathrm{Na}$, $\mathrm{K}$, $\mathrm{Cs}$, $\mathrm{Ag}$; $B = \mathrm{Nb}$). The search space included all unique supercells with determinants ranging from 1 to 8, corresponding to atomic counts from 5 to 40. This defines a pool of 49 inequivalent candidate supercells, providing a stringent test of crossover across different periodicities.

To reduce computational cost, both the exhaustive fixed-supercell baseline and the unified LGA search used the MatterSim universal potential for local relaxation and fitness evaluation. The comparison therefore probes search efficiency and supercell exploration on the same potential energy surface. The conventional baseline required 49 independent fixed-supercell evolutionary runs, one for each candidate supercell, with a population size of 10 and 10 generations per run. In contrast, Perturbation-Based LGA executed a single unified search over all 49 candidate supercells with a population size of 50 and 10 generations.

\begin{figure}[htbp]
    \centering
    \includegraphics[width=1.0\linewidth]{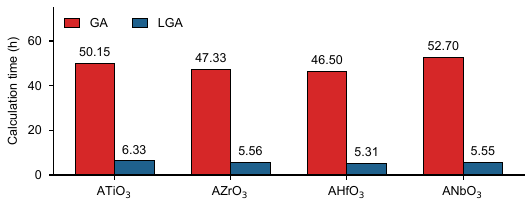}
    \caption{Unified variable-supercell LGA reduces perovskite supercell search time.
    The 16 evaluated systems are classified into four groups based on their B-site cations: ATiO$_3$, AZrO$_3$, AHfO$_3$, and ANbO$_3$. Computational search times are averaged within each group. Compared with exhaustive fixed-supercell GA searches on the same MatterSim potential energy surface, Perturbation-Based LGA achieves an approximately tenfold reduction in wall-clock search cost while recovering the same lowest-energy structures.}
    \label{fig:Pero}
\end{figure}

Across all 16 systems, the unified LGA successfully recovered the same MatterSim-ranked lowest-energy structures identified by the exhaustive fixed-supercell baseline. Crucially, the results in Fig.~\ref{fig:Pero} demonstrate a nearly order-of-magnitude reduction in search cost. This acceleration stems from replacing 49 independent fixed-supercell searches with a single variable-supercell search in which supercell topology and internal distortions evolve as coupled hereditary degrees of freedom. In contrast, direct real-space cut-and-splice operators struggle to naturally accommodate such coupled crossover.

\begin{figure*}[htbp]
    \centering

    \includegraphics[width=1.0\linewidth]{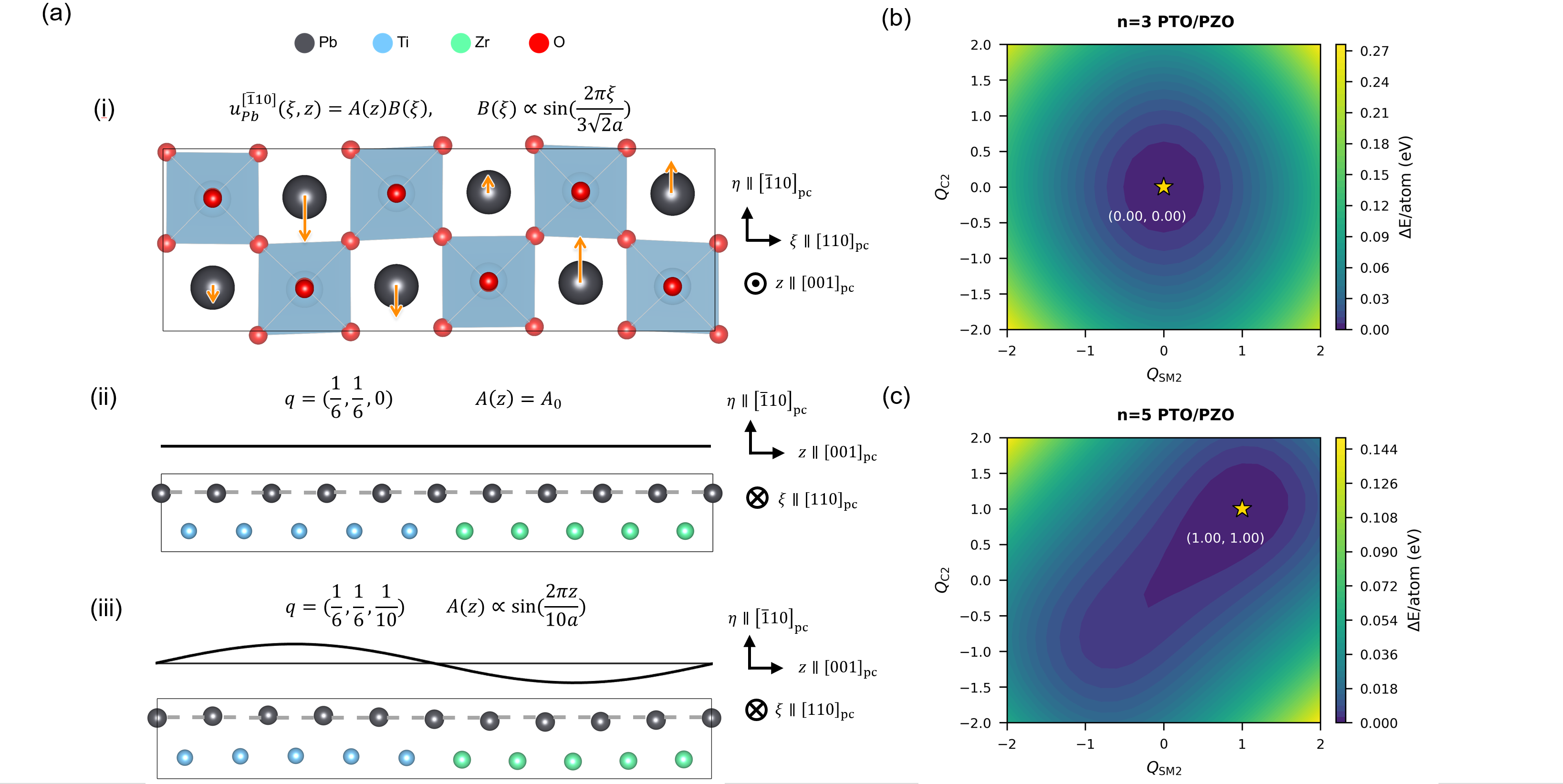}

    \caption{Thickness-dependent finite-\(q\) channel and frozen-mode energetics in PbTiO$_3$/PbZrO$_3$ superlattices. (a) Carrier-envelope representation of the additional long-period distortion modes in the \(n=5\) long-period structure. Subpanels (i)--(iii) show the common in-plane carrier, the \(q=(1/6,1/6,0)\) SM2 component, and the layer-coupled \(q=(1/6,1/6,1/10)\) C2 component, respectively. Panels (b) and (c) show frozen-mode energy maps in the normalized \((Q_{\mathrm{SM2}},Q_{\mathrm{C2}})\) space for the \(n=3\) and \(n=5\) PbTiO$_3$/PbZrO$_3$ superlattices, respectively. The \(n=3\) map shows a zero-amplitude low-energy center, whereas the \(n=5\) map shows a finite-amplitude SM2/C2 low-energy region, consistent with a thickness-dependent finite-\(q\) distortion channel. Yellow stars mark the lowest points within each map, and energies are shifted relative to the minimum in each map. The subscript \(\mathrm{pc}\) denotes pseudocubic.}
    \label{fig:distortion}
\end{figure*}

\section{Discovery of long-period phases in P\lowercase{b}T\lowercase{i}O$_3$/P\lowercase{b}Z\lowercase{r}O$_3$ superlattices} 
\label{superlattice}\label{superlattice}

As a prototypical ferroelectric perovskite, the disordered solid solution Pb(Zr$_{1-x}$Ti$_x$)O$_3$ (PZT) exhibits strong phase competition near its morphotropic phase boundary, giving rise to exceptional piezoelectric and electromechanical responses~\cite{PZT-1,PZT-2,PZT-3,PZT-4}. Consequently, PZT has become both a benchmark system for ferroic phase physics and a widely used platform for piezoelectric sensing, actuation, and MEMS technologies~\cite{PZT-5}. This complexity motivates the study of ordered PbTiO$_3$/PbZrO$_3$ superlattices, which combine ferroelectric PbTiO$_3$~\cite{PTO} with antiferroelectric PbZrO$_3$~\cite{PZO-pbam} in a layered geometry. Therefore, we investigated (001) $(\mathrm{PbTiO}_3)_n/\allowbreak(\mathrm{PbZrO}_3)_n$ superlattices with $n=1$--5 using Perturbation-Based LGA.

To manage computational costs, primary evolutionary searches for \(n=1\)--5 were driven by the MatterSim universal potential and benchmarked against independent VASP-driven searches for the computationally tractable \(n=1\) and 2 cases. All selected low-energy candidates were subsequently fully reoptimized using VASP for final energetic rankings. For \(n=1\) and 2, both MatterSim- and VASP-driven LGA converged to the same polar monoclinic \textit{Pc} symmetry with an in-plane $\sqrt{2}\times\sqrt{2}\times1$ periodicity~\cite{SL-Pc}. This agreement supports the use of MatterSim as a candidate-generation engine for larger superlattices.

For $n=3$, the VASP-refined structure retains the conventional $\sqrt{2}\times\sqrt{2}\times1$ compact periodicity. Crucially, however, for $n=4$ and $5$, LGA uncovers unusual $\sqrt{2}\times3\sqrt{2}\times1$ long-period structures that are slightly lower in energy than their compact-cell counterparts by approximately 0.5 meV atom$^{-1}$, indicating a soft competing finite-\(q\) basin. Symmetry-mode decomposition via ISODISTORT~\cite{ISODISPLACE, ISODISTORT} reveals that these long-period phases acquire additional symmetry components enabled by the enlarged in-plane periodicity, specifically, a pronounced in-plane tripling modulation with wave vector $q=(1/6,1/6,0)$ (denoted as the SM2 component). This primary modulation couples to the chemical layering along the growth direction, yielding layer-coupled sidebands (denoted as the C2 component) at $q=(1/6,1/6,1/8)$ for $n=4$ and $q=(1/6,1/6,1/10)$ for $n=5$. Distortion modes for $n=5$ are shown in Fig.~\ref{fig:distortion}(a).

To characterize the role of these symmetry components, we performed frozen-mode energy scans in the two-dimensional space spanned by the SM2 and C2 mode amplitudes. For \(n=1,2,3\), the energy minimum remains at \((Q_{\mathrm{SM2}},Q_{\mathrm{C2}})=(0,0)\), and full relaxations initialized from the unit-amplitude point relax back toward the zero-amplitude basin. For $n=4,5$, the maps show low-energy regions at finite, nearly equal SM2 and C2 amplitudes, consistent with these modes being the dominant finite-\(q\) components of the long-period basin. Representative two-dimensional frozen-mode contour maps are provided in Fig.~\ref{fig:distortion}(b) and Fig.~\ref{fig:distortion}(c). The local polar components of the \(n=4\) and \(n=5\) long-period structures both lie within the \((1\bar{1}0)_{\mathrm{pc}}\) plane, where the subscript \(\mathrm{pc}\) denotes pseudocubic. The corresponding polar axes tilt away from the \([110]_{\mathrm{pc}}\) axis by approximately \(21.6^\circ\) and \(20.7^\circ\), respectively. 

As a chemical control, we performed the same frozen-mode scan for the \((\mathrm{BaTiO}_3)_4/(\mathrm{BaZrO}_3)_4\) superlattice. The energy minimum remains at \((Q_{\mathrm{SM2}},Q_{\mathrm{C2}})=(0,0)\), and no comparable finite-amplitude SM2/C2 low-energy region appears. This comparison suggests that the long-period SM2/C2 channel is not simply imposed by the Ti/Zr layering, but depends on the Pb-based superlattice chemistry. 

To the best of our knowledge, this is the first report of low-energy in-plane long-period structures in perovskite superlattices. Previous related first-principles studies have typically used compact in-plane cells when identifying monodomain structural ground states, such as \(\sqrt{2}\times\sqrt{2}\) or \(2\times2\) cells in PbTiO$_3$/SrTiO$_3$ superlattices~\cite{PTOSTO-1,PTOSTO-2} and \(1\times1\) cells in BaTiO$_3$/SrTiO$_3$ superlattices~\cite{BTOSTO}. A possible explanation for this discovery is that the bulk PbZrO$_3$ energy landscape is unusually delicate, with various structural and polar distortions competing on a very small energy scale~\cite{PZO-80atom,PZO-antiferroelectricity,PZO-first,PZO-possibility,PZO-elastic}, thereby enabling unusual in-plane long-period structures. This application therefore demonstrates that LGA can expose hidden long-period structural order that is difficult to access with conventional fixed-periodicity or real-space GA searches.

\section{Discussion} \label{discussion}

Our results show that continuous latent representations of crystal structures can serve as operational coordinates for structural evolution. In LGA, crossover is performed between representations of parent crystals rather than between fragments of their real-space coordinates. The interpolated latent target retains information about the local atomic environments of the parents, and the offspring is obtained by inverse optimization of atomic positions and lattice vectors toward this target. In this way, LGA turns crossover into a latent-space operation while still producing explicit periodic crystal structures. In principle, this framework is not limited to the pretrained interatomic potentials used here and can be extended to other differentiable machine-learning structure encoders.

This perspective explains the improvement observed in the HfO$_2$ and perovskite benchmarks. The HfO$_2$ tests show that this mechanism suppresses offspring with high energies or short contacts and improves ground-state recovery across four pretrained encoders. The perovskite tests show that the same principle can be extended to Perturbation-Based LGA searches, where discrete supercell topology and continuous internal distortions must evolve together. Furthermore, the PbTiO$_3$/PbZrO$_3$ superlattice results illustrate why this representation-guided search is useful beyond algorithmic acceleration. LGA discovers unconventional ground-state structures featuring a $\sqrt{2}\times 3\sqrt{2}\times 1$ supercell with a long-period in-plane modulation, which has never been reported in any oxide perovskite superlattices. Symmetry-mode decomposition and frozen-mode maps identify SM2/C2 as the dominant finite-\(q\) components of this basin. 

The performance of LGA is compatible with recent discussions of convergent representations in pretrained universal interatomic potentials, which suggest that independently trained models can learn physically organized embeddings of atomic environments~\cite{Platonic,MIT}. Our contribution here is operational, showing that the representations in pretrained potentials can be used to perform a physical task that conventional coordinates handle poorly. The improvement observed across MACE, MatterSim, SevenNet, and UPET in the HfO$_2$ benchmark indicates that the usefulness of latent-space crossover is not tied to a single model architecture, even though the detailed geometry of each latent space may differ.

Furthermore, our approach provides a practical route for addressing the inverse mapping challenge inherent in generative materials design~\cite{CDVAE,Mattergen,DiffCSP}. Many latent generative approaches require a learned decoder to map latent variables back to periodic atomic structures, and this decoding step can generate invalid structures or require additional constraints. LGA circumvents the need for an explicit, learnable decoder. By reformulating the decoding step as an inverse optimization task, we convert an ill-posed mapping function into a robust search problem. This strategy helps convert interpolated latent targets into physically plausible candidates, providing a decoder-free bridge between abstract learned representations and physically realizable crystal structures.

Broadly speaking, LGA uses an energy-based fitness function, as is natural for crystal structure prediction. Moreover, the method can in principle be coupled to property-driven genetic algorithms. For example, one could search for structures to maximize polarization, tune the band gap, or optimize specific optical properties by coupling LGA to a property-based fitness function.

\begin{acknowledgments}
We acknowledge financial support from NSFC (No. 12188101), the National Key R\&D Program of China (No. 2022YFA1402901), Shanghai Science and Technology Program (No. 23JC1400900), the Guangdong Major Project of the Basic and Applied Basic Research (Future functional materials under extreme conditions--2021B0301030005), Shanghai Pilot Program for Basic Research---FuDan University 21TQ1400100 (23TQ017), the robotic AI-Scientist platform of Chinese Academy of Science, and New Cornerstone Science Foundation.
\end{acknowledgments}

\section*{Data Availability}

The data that support the findings of this article are
openly available~\cite{LGAData}.

\appendix

\section{Computational Details}
\label{app:computational}

\subsection{Ab initio structure optimization}

Density functional theory (DFT) calculations were performed using the Vienna Ab initio Simulation Package (VASP)~\cite{VASP1,VASP2,VASP3,VASP4}. The electron-ion interactions were described by the Projector Augmented Wave (PAW) method~\cite{PAW1,PAW2}. All VASP calculations in this work used the PBEsol~\cite{PBEsol} exchange-correlation functional, which is particularly suitable for lattice equilibrium properties and epitaxial strain effects in oxide perovskites. 

We used a hierarchical VASP workflow to balance robustness and accuracy across different stages of the benchmarks. Only the HfO$_2$ evolutionary-loop benchmark used a three-stage relaxation protocol. In this workflow, candidate structures generated by either USPEX or Random-Based LGA were sequentially relaxed with progressively tighter k-point densities and convergence criteria. The third stage of the HfO$_2$ workflow was then used as the common final VASP protocol for all VASP calculations outside the HfO$_2$ evolutionary loop. INCAR settings, k-point-generation rules and PAW datasets for the three-stage HfO$_2$ workflow are provided in Supplemental Material.

\subsection{Computational roles of VASP and pretrained interatomic potentials}

VASP and pretrained interatomic potentials were used for distinct purposes in different parts of the workflow. In the LGA crossover operator, pretrained models provide differentiable latent encoders for constructing and optimizing the target representation \( \mathbf{z}_{\mathrm{target}} \). In the HfO$_2$ benchmark, MACE, MatterSim, SevenNet, and UPET are used only as latent encoders; all offspring generated by \textsc{USPEX} and Random-Based LGA are relaxed, ranked, and assessed with VASP. In the perovskite benchmark, MatterSim defines the common potential energy surface for both the exhaustive fixed-supercell baseline and the unified search over candidate supercells. In the PbTiO$_3$/PbZrO$_3$ superlattice study, MatterSim is used for candidate generation, whereas all selected low-energy structures are reoptimized with VASP before final energetic and mode analyses.

\section{Latent Descriptors and Inverse Optimization}
\label{app:latent}

\subsection{Latent descriptor extraction}

For each universal potential, LGA uses the differentiable atomic descriptors exposed by the corresponding model-specific forward pass. These per-atom descriptors are mean-pooled over atoms to construct the global latent representation \(\mathbf{z}_X\), as defined in Eq.~\ref{Eq1}. For MACE, the descriptor is obtained from the internal \texttt{node\_feats} output. For MatterSim, SevenNet, and UPET, the hidden atomic features exposed by the corresponding model interfaces are used. No additional training, fine-tuning, or learned projection was introduced.

\subsection{Details for the optimization solver}

In our implementation, the inverse optimization task ($X^* = \operatorname*{arg\,min} \| \mathbf{z}_{X} - \mathbf{z}_{\text{target}} \|_2$) is executed using a custom ASE~\cite{ASE} calculator interface. The optimization process is driven by the Fast Inertial Relaxation Engine (FIRE) algorithm~\cite{FIRE}, employing a FrechetCellFilter to enable the simultaneous relaxation of atomic positions and lattice parameters. The optimization proceeds for a maximum of 300 steps, ensuring sufficient traversal of the latent representation space to reach the target configuration.

\subsection{Weighting strategies for latent interpolation}

As defined in Sec. ~\ref{mechanism}, the target latent vector $\mathbf{z}_{\text{target}}$ is determined by the interpolation weights $\omega_{1}$ and $\omega_{2}$, normalized such that $\omega_{1} + \omega_{2} = 1$. We implemented four distinct strategies to define these weights:

\begin{itemize}[leftmargin=*, noitemsep, topsep=0pt]
    \item \textit{Uniform Weighting}: Serves as an unbiased baseline by assigning equal contributions to both parents:
    \begin{equation}
        \omega_i = 1/2 .
    \end{equation}

    \item \textit{Boltzmann Distribution Weighting}: Inspired by statistical mechanics, this scheme adjusts selection pressure via a ``temperature'' parameter $\tau$:
    \begin{equation}
        \omega_i = \frac{e^{-E_i/\tau}}{e^{-E_1/\tau} + e^{-E_2/\tau}} .
    \end{equation}

    \item \textit{Linear Scaling}: Weights are coupled to the relative quality of individuals within the current generation. Defining the fitness as $f_i = E_{\max} - E_i$, the weight is calculated as
    \begin{equation}
        \omega_i = \frac{f_i}{f_1 + f_2} .
    \end{equation}

    \item \textit{Rank-Based Weighting}: To mitigate the influence of outliers, weights are derived from the population rank. Let $N_{pop}$ be the total population size and $R_i$ be the rank of parent $i$, where $R_{i}=1$ represents the structure with the lowest energy. Defining rank score as $S_i = N_{pop} - R_i + 1$, the weight is calculated as
    \begin{equation}
        \omega_i = \frac{S_i}{S_1+S_2} .
    \end{equation}
\end{itemize}

\section{Evolutionary Search Workflows}
\label{app:workflows}

\subsection{Random-Based Latent Genetic Algorithm}

The random-based search workflow follows the conventional GA initialization and evolution scheme. PASP generates random symmetry-constrained structures by sampling the number of atoms within a user-defined range, selecting a random space group, and constructing the corresponding configuration with its internal structure generator. The primary distinction between Random-Based LGA and traditional random-based GAs is that LGA optimizes the offspring atomic positions and lattice vectors to match the interpolated latent target instead of applying a real-space operation.

\subsection{Perturbation-Based Latent Genetic Algorithm}

Perturbation-Based Latent Genetic Algorithm enables the co-evolution of supercell topology and internal atomic configuration. By formalizing the lattice search space using Hermite Normal Forms (HNFs) and coupling it with latent-space structural refinement, the algorithm efficiently navigates the vast configuration space defined by both lattice dimensions and atomic degrees of freedom. The workflow is structured into two key components.

\appsubhead{a.\quad HNF Pool Generation}

Prior to evolution, a pool of inequivalent supercells is generated to define the search space. The candidate list is constructed through a three-step filtering process: (1) generation of all unique integer matrices in lower-triangular HNF format within a specified determinant range; (2) removal of symmetry-equivalent HNFs under the system's point group operations; and (3) removal of redundant supercells whose periodicities are already represented within another retained supercell~\cite{HNF}.

\appsubhead{b.\quad Dual-Cross Evolution}

The evolutionary cycle employs a hybrid mechanism to simultaneously optimize periodicity and geometry:

\begin{itemize}[leftmargin=*, noitemsep, topsep=0pt]
    \item \textit{Cell Crossover}. Supercell dimensions are inherited by performing column-wise cut-and-splice operations on parent HNFs. The resulting matrix is mapped to the unique HNF in the precomputed pool to ensure crystallographic validity.
    \item \textit{Structural Crossover}. Once the offspring supercell is determined, the offspring are initialized with random distortions, and subsequently optimized in the latent space to match the interpolated target vector $\mathbf{z}_{\text{target}}$.
\end{itemize}

\section{Structural Mode Analysis}
\label{app:modes}

For the analysis of structural distortion modes, the five-atom cubic \(Pm\bar{3}m\) perovskite structure was used as the reference parent phase. The lattice constant of this cubic cell is denoted by \(a\), and all wave vectors \(q\) are expressed in units of \(2\pi/a\).

The SM2 and C2 mode patterns used in Fig.~\ref{fig:distortion} were defined from ISODISTORT decompositions of the VASP-relaxed long-period structures. The coordinates $Q_{\mathrm{SM2}}$ and $Q_{\mathrm{C2}}$ denote normalized amplitudes of these two components. Frozen-mode structures were generated by linearly combining the normalized SM2 and C2 distortion patterns, and the resulting static energies were evaluated using the same VASP protocol.

\bibliography{cite}

@book{CSP1,
  title={Modern methods of crystal structure prediction},
  author={Oganov, Artem R},
  year={2011},
  publisher={John Wiley \& Sons}
}

@article{CSP2,
  title={Crystal structure prediction from first principles},
  author={Woodley, Scott M and Catlow, Richard},
  journal={Nature Materials},
  volume={7},
  number={12},
  pages={937--946},
  year={2008},
  publisher={Nature Publishing Group UK London}
}

@book{GA,
  title={Adaptation in natural and artificial systems: an introductory analysis with applications to biology, control, and artificial intelligence},
  author={Holland, John H},
  year={1992},
  publisher={MIT Press}
}

@article{USPEX1,
  title={Crystal structure prediction using ab initio evolutionary techniques: Principles and applications},
  author={Oganov, Artem R and Glass, Colin W},
  journal={The Journal of chemical physics},
  volume={124},
  number={24},
  year={2006},
  publisher={AIP Publishing}
}

@article{USPEX2,
  title={How Evolutionary Crystal Structure Prediction Works--and Why},
  author={Oganov, Artem R and Lyakhov, Andriy O and Valle, Mario},
  journal={Accounts of chemical research},
  volume={44},
  number={3},
  pages={227--237},
  year={2011},
  publisher={ACS Publications}
}

@article{USPEX3,
  title={New developments in evolutionary structure prediction algorithm {USPEX}},
  author={Lyakhov, Andriy O and Oganov, Artem R and Stokes, Harold T and Zhu, Qiang},
  journal={Computer Physics Communications},
  volume={184},
  number={4},
  pages={1172--1182},
  year={2013},
  publisher={Elsevier}
}

@article{CALYPSO,
  title={{CALYPSO}: A method for crystal structure prediction},
  author={Wang, Yanchao and Lv, Jian and Zhu, Li and Ma, Yanming},
  journal={Computer Physics Communications},
  volume={183},
  number={10},
  pages={2063--2070},
  year={2012},
  publisher={Elsevier}
}

@article{AGA,
  title={An adaptive genetic algorithm for crystal structure prediction},
  author={Wu, SQ and Ji, Min and Wang, Cai-Zhuang and Nguyen, Manh Cuong and Zhao, Xin and Umemoto, K and Wentzcovitch, RM and Ho, Kai-Ming},
  journal={Journal of Physics: Condensed Matter},
  volume={26},
  number={3},
  pages={035402},
  year={2013},
  publisher={IOP Publishing}
}

@article{MAGUS1,
  title={{MAGUS}: machine learning and graph theory assisted universal structure searcher},
  author={Wang, Junjie and Gao, Hao and Han, Yu and Ding, Chi and Pan, Shuning and Wang, Yong and Jia, Qiuhan and Wang, Hui-Tian and Xing, Dingyu and Sun, Jian},
  journal={National Science Review},
  volume={10},
  number={7},
  pages={nwad128},
  year={2023},
  publisher={Oxford University Press}
}

@article{Xtalopt,
  title={{XtalOpt}: An open-source evolutionary algorithm for crystal structure prediction},
  author={Lonie, David C and Zurek, Eva},
  journal={Computer Physics Communications},
  volume={182},
  number={2},
  pages={372--387},
  year={2011},
  publisher={Elsevier}
}

@article{EVO,
  title={{EVO}—Evolutionary algorithm for crystal structure prediction},
  author={Bahmann, Silvia and Kortus, Jens},
  journal={Computer Physics Communications},
  volume={184},
  number={6},
  pages={1618--1625},
  year={2013},
  publisher={Elsevier}
}

@article{AIRSS,
  title={Ab initio random structure searching},
  author={Pickard, Chris J and Needs, RJ},
  journal={Journal of Physics: Condensed Matter},
  volume={23},
  number={5},
  pages={053201},
  year={2011},
  publisher={IOP Publishing}
}

@article{Hopping,
  title={Minima hopping: An efficient search method for the global minimum of the potential energy surface of complex molecular systems},
  author={Goedecker, Stefan},
  journal={The Journal of Chemical Physics},
  volume={120},
  number={21},
  pages={9911--9917},
  year={2004},
  publisher={American Institute of Physics}
}

@article{Case-Xtal,
  title={Pressure-stabilized sodium polyhydrides: {NaH$_n$} ($n>1$)},
  author={Baettig, Pio and Zurek, Eva},
  journal={Physical Review Letters},
  volume={106},
  number={23},
  pages={237002},
  year={2011},
  publisher={APS}
}

@article{Case-EVO,
  title={Metastable structure of {Li$_{13}$Si$_4$}},
  author={Gruber, Thomas and Bahmann, Silvia and Kortus, Jens},
  journal={Physical Review B},
  volume={93},
  number={14},
  pages={144104},
  year={2016},
  publisher={APS}
}

@article{Case-AGA,
  title={Exploring the structural complexity of intermetallic compounds by an adaptive genetic algorithm},
  author={Zhao, X and Nguyen, MC and Zhang, WY and Wang, CZ and Kramer, Matthew J and Sellmyer, David J and Li, XZ and Zhang, F and Ke, LQ and Antropov, Vladimir P and others},
  journal={Physical Review Letters},
  volume={112},
  number={4},
  pages={045502},
  year={2014},
  publisher={APS}
}

@article{Case-Caylpso,
  title={Reactions of xenon with iron and nickel are predicted in the Earth's inner core},
  author={Zhu, Li and Liu, Hanyu and Pickard, Chris J and Zou, Guangtian and Ma, Yanming},
  journal={Nature Chemistry},
  volume={6},
  number={7},
  pages={644--648},
  year={2014},
  publisher={Nature Publishing Group UK London}
}

@article{Case-MAGUS,
  title={Mixed coordination silica at megabar pressure},
  author={Liu, Cong and Shi, Jiuyang and Gao, Hao and Wang, Junjie and Han, Yu and Lu, Xiancai and Wang, Hui-Tian and Xing, Dingyu and Sun, Jian},
  journal={Physical Review Letters},
  volume={126},
  number={3},
  pages={035701},
  year={2021},
  publisher={APS}
}

@article{Case-PASP1,
  title={Prediction of two-dimensional materials by the global optimization approach},
  author={Gu, Teng and Luo, Wei and Xiang, Hongjun},
  journal={Wiley Interdisciplinary Reviews: Computational Molecular Science},
  volume={7},
  number={2},
  pages={e1295},
  year={2017},
  publisher={Wiley Online Library}
}

@article{Case-PASP2,
  title={Genetic algorithm prediction of pressure-induced multiferroicity in the perovskite {PbCoO$_3$}},
  author={Lou, Feng and Luo, Wei and Feng, Junsheng and Xiang, Hongjun},
  journal={Physical Review B},
  volume={99},
  number={20},
  pages={205104},
  year={2019},
  publisher={APS}
}

@article{Behler,
  title={Generalized neural-network representation of high-dimensional potential-energy surfaces},
  author={Behler, J{\"o}rg and Parrinello, Michele},
  journal={Physical Review Letters},
  volume={98},
  number={14},
  pages={146401},
  year={2007},
  publisher={APS}
}

@article{MACE,
  title={{MACE}: Higher order equivariant message passing neural networks for fast and accurate force fields},
  author={Batatia, Ilyes and Kovacs, David P and Simm, Gregor and Ortner, Christoph and Cs{\'a}nyi, G{\'a}bor},
  journal={Advances in Neural Information Processing Systems},
  volume={35},
  pages={11423--11436},
  year={2022}
}

@article{Seven-omni,
  title   = {Optimizing cross-domain transfer for universal machine learning interatomic potentials},
  author  = {Kim, Jaesun and You, Jinmu and Park, Yutack and others},
  journal = {Nature Communications},
  volume  = {17},
  pages   = {3432},
  year    = {2026},
  doi     = {10.1038/s41467-026-70195-8}
}

@article{Sevennet,
	title = {Scalable Parallel Algorithm for Graph Neural Network Interatomic Potentials in Molecular Dynamics Simulations},
	volume = {20},
	doi = {10.1021/acs.jctc.4c00190},
	number = {11},
	journal = {J. Chem. Theory Comput.},
	author = {Park, Yutack and Kim, Jaesun and Hwang, Seungwoo and Han, Seungwu},
	year = {2024},
	pages = {4857--4868},
}

@article{M3GNET,
  title={A universal graph deep learning interatomic potential for the periodic table},
  author={Chen, Chi and Ong, Shyue Ping},
  journal={Nature Computational Science},
  volume={2},
  number={11},
  pages={718--728},
  year={2022},
  publisher={Nature Publishing Group US New York}
}

@article{CHGNET,
  title={{CHGNet} as a pretrained universal neural network potential for charge-informed atomistic modelling},
  author={Deng, Bowen and Zhong, Peichen and Jun, KyuJung and Riebesell, Janosh and Han, Kevin and Bartel, Christopher J and Ceder, Gerbrand},
  journal={Nature Machine Intelligence},
  volume={5},
  number={9},
  pages={1031--1041},
  year={2023},
  publisher={Nature Publishing Group UK London}
}

@article{Mattersim,
      title={{MatterSim}: A Deep Learning Atomistic Model Across Elements, Temperatures and Pressures},
      author={Han Yang and Chenxi Hu and Yichi Zhou and Xixian Liu and Yu Shi and Jielan Li and Guanzhi Li and Zekun Chen and Shuizhou Chen and Claudio Zeni and Matthew Horton and Robert Pinsler and Andrew Fowler and Daniel Zügner and Tian Xie and Jake Smith and Lixin Sun and Qian Wang and Lingyu Kong and Chang Liu and Hongxia Hao and Ziheng Lu},
      year={2024},
      eprint={2405.04967},
      archivePrefix={arXiv},
      primaryClass={cond-mat.mtrl-sci},
      url={https://arxiv.org/abs/2405.04967},
      journal={arXiv preprint arXiv:2405.04967}
}

@article{UPET-MAD,
  title={{PET-MAD} as a lightweight universal interatomic potential for advanced materials modeling},
  author={Mazitov, Arslan and Bigi, Filippo and Kellner, Matthias and Pegolo, Paolo and Tisi, Davide and Fraux, Guillaume and Pozdnyakov, Sergey and Loche, Philip and Ceriotti, Michele},
  journal={Nature Communications},
  volume={16},
  number={1},
  pages={10653},
  year={2025},
  publisher={Nature Publishing Group UK London}
}

@article{UPET,
  title = {Smooth, Exact Rotational Symmetrization for Deep Learning on Point Clouds},
  journal = {Advances in {{Neural Information Processing Systems}}},
  author = {Pozdnyakov, Sergey and Ceriotti, Michele},
  year = 2023,
  volume = {36},
  pages = {79469--79501},
}

@article{Platonic,
  author  = {Li, Z. and Walsh, A.},
  title   = {Platonic representation of foundation machine learning interatomic potentials},
  journal = {Nature Machine Intelligence},
  volume  = {8},
  pages   = {830--840},
  year    = {2026},
  doi     = {10.1038/s42256-026-01235-7},
  url     = {https://doi.org/10.1038/s42256-026-01235-7}
}

@article{PASP,
  title={{PASP}: Property analysis and simulation package for materials},
  author={Lou, Feng and Li, XY and Ji, JY and Yu, HY and Feng, JS and Gong, XG and Xiang, HJ},
  journal={The Journal of Chemical Physics},
  volume={154},
  number={11},
  pages={114103},
  year={2021},
  publisher={AIP Publishing}
}

@article{HNF,
  title={Algorithm for generating derivative structures},
  author={Hart, Gus LW and Forcade, Rodney W},
  journal={Physical Review B—Condensed Matter and Materials Physics},
  volume={77},
  number={22},
  pages={224115},
  year={2008},
  publisher={APS}
}

@article{MIT,
  title={Universally converging representations of matter across scientific foundation models},
  author={Edamadaka, Sathya and Yang, Soojung and Li, Ju and G{\'o}mez-Bombarelli, Rafael},
  journal={arXiv preprint arXiv:2512.03750},
  year={2025}
}

@article{FIRE,
  title={Structural relaxation made simple},
  author={Bitzek, Erik and Koskinen, Pekka and G{\"a}hler, Franz and Moseler, Michael and Gumbsch, Peter},
  journal={Physical Review Letters},
  volume={97},
  number={17},
  pages={170201},
  year={2006},
  publisher={APS}
}

@article{ASE,
  title={The atomic simulation environment—a Python library for working with atoms},
  author={Larsen, Ask Hjorth and Mortensen, Jens J{\o}rgen and Blomqvist, Jakob and Castelli, Ivano E and Christensen, Rune and Du{\l}ak, Marcin and Friis, Jesper and Groves, Michael N and Hammer, Bj{\o}rk and Hargus, Cory and others},
  journal={Journal of Physics: Condensed Matter},
  volume={29},
  number={27},
  pages={273002},
  year={2017},
  publisher={IOP Publishing}
}

@article{VASP1,
  title={Ab initio molecular dynamics for liquid metals},
  author={Kresse, Georg and Hafner, J{\"u}rgen},
  journal={Physical Review B},
  volume={47},
  number={1},
  pages={558},
  year={1993},
  publisher={APS}
}

@article{VASP2,
  title={Ab initio molecular-dynamics simulation of the liquid-metal--amorphous-semiconductor transition in germanium},
  author={Kresse, Georg and Hafner, J{\"u}rgen},
  journal={Physical Review B},
  volume={49},
  number={20},
  pages={14251},
  year={1994},
  publisher={APS}
}

@article{VASP3,
  title={Efficiency of ab-initio total energy calculations for metals and semiconductors using a plane-wave basis set},
  author={Kresse, Georg and Furthm{\"u}ller, J{\"u}rgen},
  journal={Computational Materials Science},
  volume={6},
  number={1},
  pages={15--50},
  year={1996},
  publisher={Elsevier}
}

@article{VASP4,
  title={Efficient iterative schemes for ab initio total-energy calculations using a plane-wave basis set},
  author={Kresse, Georg and Furthm{\"u}ller, J{\"u}rgen},
  journal={Physical Review B},
  volume={54},
  number={16},
  pages={11169},
  year={1996},
  publisher={APS}
}

@article{PAW1,
  title={Projector augmented-wave method},
  author={Bl{\"o}chl, Peter E},
  journal={Physical Review B},
  volume={50},
  number={24},
  pages={17953},
  year={1994},
  publisher={APS}
}

@article{PAW2,
  title={From ultrasoft pseudopotentials to the projector augmented-wave method},
  author={Kresse, Georg and Joubert, Daniel},
  journal={Physical Review B},
  volume={59},
  number={3},
  pages={1758},
  year={1999},
  publisher={APS}
}

@article{PBEsol,
  title={Restoring the density-gradient expansion for exchange in solids and surfaces},
  author={Perdew, John P and Ruzsinszky, Adrienn and Csonka, G{\'a}bor I and Vydrov, Oleg A and Scuseria, Gustavo E and Constantin, Lucian A and Zhou, Xiaolan and Burke, Kieron},
  journal={Physical Review Letters},
  volume={100},
  number={13},
  pages={136406},
  year={2008},
  publisher={APS}
}

@article{SL-Pc,
  title={Ground state and properties of ferroelectric superlattices based on crystals of the perovskite family},
  author={Lebedev, AI},
  journal={Physics of the Solid State},
  volume={52},
  number={7},
  pages={1448--1462},
  year={2010},
  publisher={Springer}
}

@article{CDVAE,
  title={Crystal diffusion variational autoencoder for periodic material generation},
  author={Xie, Tian and Fu, Xiang and Ganea, Octavian-Eugen and Barzilay, Regina and Jaakkola, Tommi},
  journal={arXiv preprint arXiv:2110.06197},
  year={2021}
}

@article{DiffCSP,
  title={Crystal structure prediction by joint equivariant diffusion},
  author={Jiao, Rui and Huang, Wenbing and Lin, Peijia and Han, Jiaqi and Chen, Pin and Lu, Yutong and Liu, Yang},
  journal={Advances in Neural Information Processing Systems},
  volume={36},
  pages={17464--17497},
  year={2023}
}

@article{Mattergen,
  title={A generative model for inorganic materials design},
  author={Zeni, Claudio and Pinsler, Robert and Z{\"u}gner, Daniel and Fowler, Andrew and Horton, Matthew and Fu, Xiang and Wang, Zilong and Shysheya, Aliaksandra and Crabb{\'e}, Jonathan and Ueda, Shoko and others},
  journal={Nature},
  volume={639},
  number={8055},
  pages={624--632},
  year={2025},
  publisher={Nature Publishing Group UK London}
}

@article{PZT-1,
  author  = {Noheda, B. and Cox, D. E. and Shirane, G. and Gonzalo, J. A. and Cross, L. E. and Park, S.-E.},
  title   = {A monoclinic ferroelectric phase in the {Pb(Zr$_{1-x}$Ti$_x$)O$_3$} solid solution},
  journal = {Applied Physics Letters},
  volume  = {74},
  pages   = {2059--2061},
  year    = {1999},
  doi     = {10.1063/1.123756}
}

@article{PZT-2,
  author  = {Bellaiche, L. and Garc{\'i}a, A. and Vanderbilt, D.},
  title   = {Finite-temperature properties of {Pb(Zr$_{1-x}$Ti$_x$)O$_3$} alloys from first principles},
  journal = {Physical Review Letters},
  volume  = {84},
  pages   = {5427--5430},
  year    = {2000},
  doi     = {10.1103/PhysRevLett.84.5427}
}

@article{PZT-3,
  author  = {Guo, R. and Cross, L. E. and Park, S.-E. and Noheda, B. and Cox, D. E. and Shirane, G.},
  title   = {Origin of the High Piezoelectric Response in {PbZr$_{1-x}$Ti$_x$O$_3$}},
  journal = {Physical Review Letters},
  volume  = {84},
  pages   = {5423--5426},
  year    = {2000},
  doi     = {10.1103/PhysRevLett.84.5423}
}

@article{PZT-4,
  author  = {Damjanovic, D.},
  title   = {Contributions to the piezoelectric effect in ferroelectric single crystals and ceramics},
  journal = {Journal of the American Ceramic Society},
  volume  = {88},
  pages   = {2663--2676},
  year    = {2005},
  doi     = {10.1111/j.1551-2916.2005.00671.x}
}

@article{PZT-5,
  author  = {Muralt, P. and Polcawich, R. G. and Trolier-McKinstry, S.},
  title   = {Piezoelectric thin films for sensors, actuators, and energy harvesting},
  journal = {MRS Bulletin},
  volume  = {34},
  number  = {9},
  pages   = {658--664},
  year    = {2009},
  doi     = {10.1557/mrs2009.177}
}

@article{PTO,
  author  = {Shirane, Gen and Hoshino, Sadao and Suzuki, Kazuo},
  title   = {X-Ray Study of the Phase Transition in Lead Titanate},
  journal = {Physical Review},
  volume  = {80},
  pages   = {1105--1106},
  year    = {1950},
  doi     = {10.1103/PhysRev.80.1105}
}

@article{PZO-possibility,
  title={On the possibility that {PbZrO$_3$} not be antiferroelectric},
  author={Aramberri, Hugo and Cazorla, Claudio and Stengel, Massimiliano and {\'I}{\~n}iguez, Jorge},
  journal={npj Computational Materials},
  volume={7},
  number={1},
  pages={196},
  year={2021},
  publisher={Nature Publishing Group UK London}
}

@article{PZO-elastic,
  title={First-principles investigations of elastic properties and energetics of antiferroelectric and ferroelectric phases of {PbZrO$_3$}},
  author={Kagimura, Ricardo and Singh, David J},
  journal={Physical Review B—Condensed Matter and Materials Physics},
  volume={77},
  number={10},
  pages={104113},
  year={2008},
  publisher={APS}
}

@misc{PZO-80atom,
      title={A re-examination of antiferroelectric {PbZrO$_3$} and {PbHfO$_3$}: an 80-atom {$Pnam$} structure}, 
      author={J. S. Baker and M. Paściak and J. K. Shenton and P. Vales-Castro and B. Xu and J. Hlinka and P. Márton and R. G. Burkovsky and G. Catalan and A. M. Glazer and D. R. Bowler},
      year={2021},
      eprint={2102.08856},
      archivePrefix={arXiv},
      primaryClass={cond-mat.mtrl-sci},
      url={https://arxiv.org/abs/2102.08856}, 
}

@article{PZO-pbam,
  title={Antiferroelectric structure of lead zirconate},
  author={Sawaguchi, Etsuro and Maniwa, H and Hoshino, Sadao},
  journal={Physical Review},
  volume={83},
  number={5},
  pages={1078},
  year={1951},
  publisher={APS}
}

@article{PZO-antiferroelectricity,
  title={Antiferroelectricity and ferroelectricity in epitaxially strained {PbZrO$_3$} from first principles},
  author={Reyes-Lillo, Sebastian E and Rabe, Karin M},
  journal={Physical Review B—Condensed Matter and Materials Physics},
  volume={88},
  number={18},
  pages={180102},
  year={2013},
  publisher={APS}
}

@article{PZO-first,
  title={First-principles study of the multimode antiferroelectric transition in {PbZrO$_3$}},
  author={{\'I}{\~n}iguez, Jorge and Stengel, Massimiliano and Prosandeev, Sergey and Bellaiche, L},
  journal={Physical Review B},
  volume={90},
  number={22},
  pages={220103},
  year={2014},
  publisher={APS}
}

@article{PTOSTO-2,
  author  = {Aguado-Puente, P. and Garc{\'i}a-Fern{\'a}ndez, P. and Junquera, J.},
  title   = {Interplay of couplings between antiferrodistortive, ferroelectric, and strain degrees of freedom in monodomain {PbTiO$_3$/SrTiO$_3$} superlattices},
  journal = {Physical Review Letters},
  volume  = {107},
  pages   = {217601},
  year    = {2011},
  doi     = {10.1103/PhysRevLett.107.217601}
}

@article{PTOSTO-1,
  title   = {Determination of ground-state and low-energy structures of perovskite superlattices from first principles},
  author  = {Zhou, Yuanjun and Rabe, Karin M.},
  journal = {Physical Review B},
  volume  = {89},
  pages   = {214108},
  year    = {2014},
  doi     = {10.1103/PhysRevB.89.214108}
}

@article{BTOSTO,
  title={Dielectric, piezoelectric, and elastic properties of {BaTiO$_3$/SrTiO$_3$} ferroelectric superlattices from first principles},
  author={Lebedev, Alexander I},
  journal={Journal of Advanced Dielectrics},
  volume={2},
  number={01},
  pages={1250003},
  year={2012},
  publisher={World Scientific}
}

@article{ISODISPLACE,
  title={{ISODISPLACE}: a web-based tool for exploring structural distortions},
  author={Campbell, Branton J and Stokes, Harold T and Tanner, David E and Hatch, Dorian M},
  journal={Journal of Applied Crystallography},
  volume={39},
  number={4},
  pages={607--614},
  year={2006},
  publisher={International Union of Crystallography}
}

@misc{ISODISTORT,
  author       = {Stokes, H. T. and Hatch, D. M. and Campbell, B. J.},
  title        = {{ISODISTORT, ISOTROPY Software Suite}},
  howpublished = {\url{https://iso.byu.edu}}
}

@misc{LGAData,
  author       = {Zheng, Kaixin and Yin, Wanjian and Yu, Hongyu and Xiang, Hongjun},
  title        = {{LGA data repository}},
  year         = {2026},
  howpublished = {\url{https://github.com/KxZhenggg/LGA}}
}

\end{document}

\title{%
{\normalfont\normalsize Supplemental Material}\\[2.0em]
{\bfseries Latent Genetic Algorithm for Crystal Structure Prediction}
}

\author{Kaixin Zheng}
\affiliation{Key Laboratory of Computational Physical Sciences (Ministry of Education), Institute of Computational Physical Sciences, State Key Laboratory of Surface Physics, and Department of Physics, Fudan University, Shanghai 200433, China}

\author{Wanjian Yin}
\affiliation{College of Energy, Soochow Institute for Energy and Materials Innovations, Soochow University, Suzhou 215006, China}

\author{Hongyu Yu}
\email[Contact author: ]{hongyuyu20@fudan.edu.cn}
\affiliation{Key Laboratory of Computational Physical Sciences (Ministry of Education), Institute of Computational Physical Sciences, State Key Laboratory of Surface Physics, and Department of Physics, Fudan University, Shanghai 200433, China}

\author{Hongjun Xiang}
\email[Contact author: ]{hxiang@fudan.edu.cn}
\affiliation{Key Laboratory of Computational Physical Sciences (Ministry of Education), Institute of Computational Physical Sciences, State Key Laboratory of Surface Physics, and Department of Physics, Fudan University, Shanghai 200433, China}

\maketitle
\thispagestyle{empty}

\clearpage

For the VASP calculations performed in this work, we used the following VASP PBE PAW datasets and explicitly treated valence configurations: \texttt{Pb} (\(5d^{10}6s^26p^2\)), \texttt{Ba\_sv} (\(5s^25p^66s^2\)), \texttt{Ti\_sv} (\(3s^23p^63d^24s^2\)), \texttt{Zr\_sv} (\(4s^24p^64d^25s^2\)), \texttt{Hf\_pv} (\(5p^65d^26s^2\)), and \texttt{O} (\(2s^22p^4\)). Detailed parameters are shown in Table~\ref{tab:vasp_parameters}. For the HfO$_2$ benchmark, the target $k$-point densities were matched between the PASP/LGA and USPEX workflows. Small differences in the resulting integer meshes arose from code-specific rounding in automatic mesh generation, with the USPEX meshes being identical to or marginally denser than the corresponding VASP meshes. Final structural relaxations and success assessments used the same stage-III VASP protocol.

\begin{table}
\caption{\label{tab:vasp_parameters}
VASP calculation parameters used in this work. The HfO$_2$ evolutionary loop used a three-stage relaxation workflow, whereas the stage-III settings were used as the common final VASP protocol for all VASP calculations outside the HfO$_2$ evolutionary loop.}
\begin{ruledtabular}
\begin{tabular}{lccc}
Parameter & Stage I & Stage II & Stage III \\
\colrule
ENCUT & 550 eV & 550 eV & 550 eV \\
PREC & Normal & Normal & Accurate \\
K-point setting & automatic length = 5 & automatic length = 10 & automatic length = 15 \\
USPEX KresolStart & 0.2 & 0.1 & 0.0667 \\
EDIFF & \(1 \times 10^{-4}\) eV & \(1 \times 10^{-5}\) eV & \(1 \times 10^{-6}\) eV \\
EDIFFG & \(-2 \times 10^{-1}\) eV \AA\(^{-1}\) & \(-1 \times 10^{-1}\) eV \AA\(^{-1}\) & \(-1 \times 10^{-2}\) eV \AA\(^{-1}\) \\
\end{tabular}
\end{ruledtabular}
\end{table}

\begin{figure}[t]
    \centering
    \includegraphics[width=\textwidth]{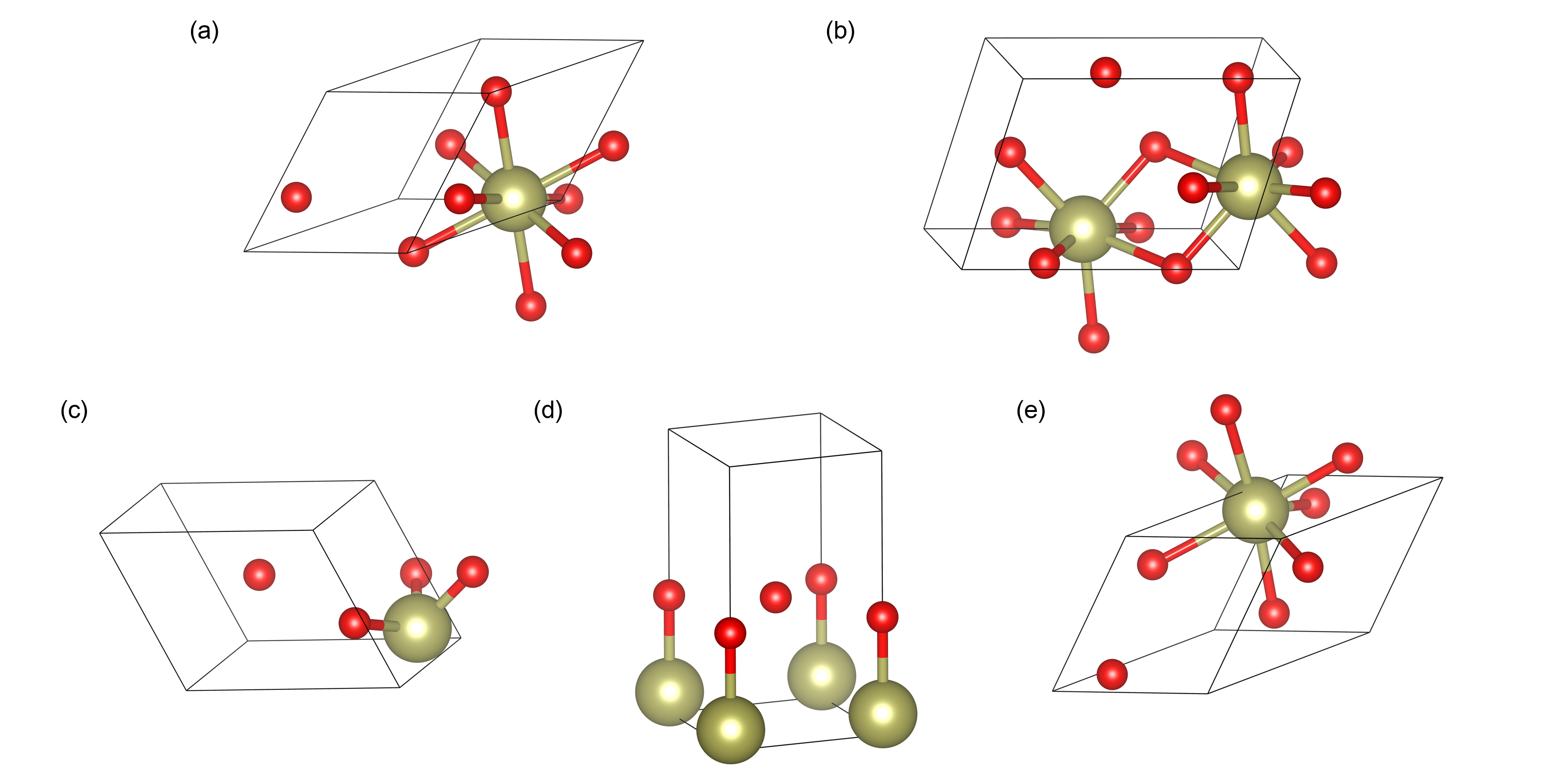}
    \caption{\label{fig:crystal_fig2}
    Structural visualizations of the key structures shown in Fig. 2 of the main text. (a) and (b) are the parent structures. (c) represents the offspring generated by the traditional GA. (d) and (e) denote the initial and final structure of the latent-space crossover, respectively.}
\end{figure}